\documentclass[twocolumn, prb, aps, showpacs, floatfix, letterpaper]{revtex4}

\usepackage[dvips]{graphicx} % Include figure files
\usepackage{latexsym}
\usepackage{dcolumn} % Align table columns on decimal point
\usepackage{bm} % bold math
\usepackage{textcomp} %for special symbols

\begin{document}
\title{
Pressure induced reentrant electronic and magnetic state in Pr$_{0.7}$Ca$_{0.3}$MnO$_{3}$ manganite
}

\author{Congwu Cui}
\author{Trevor A. Tyson}
\affiliation{Physics Department, New Jersey Institute of Technology, Newark, New Jersey 07102}

\date{received 10 June 2003}

\begin{abstract}

In Pr$_{0.7}$Ca$_{0.3}$MnO$_{3}$, pressure induces reentrant magnetic and electronic state changes in the range 1 atm to $\sim$ 6 GPa. The metal-insulator and magnetic transition temperatures coincide from $\sim$1 to 5 GPa, decouple outside of this range and do not change monotonically with pressure. The effects may be explained by pressure tuned competition between double exchange and super exchange. The insulating state induced by pressure above $\sim$5 GPa is possibly ferromagnetic, different from the ferromagnetic and antiferromagnetic phase-separated insulating state below $\sim$0.8 GPa.

\end{abstract}

%PACS numbers: 

\maketitle

The colossal magnetoresistive manganite Pr$_{1-x}$Ca$_{x}$MnO$_{3}$ is a narrow bandwidth system due to the large mismatch between the Pr/Ca and Mn ion sizes. In Pr$_{1-x}$Ca$_{x}$MnO$_{3}$, it has been found that magnetic fields,\cite{tomioka_prb_53_1689_96, barratt_apl_68_424_96} high electric fields,\cite{asamitsu_nature_388_59_97} irradiation with x-rays,\cite{kiryukhin_nature_386_813_97, cox_prb_57_3305_98} electrons,\cite{hervieu_prb_60_726_99} or visible light\cite{miyano_prl_78_4257_97, mori_jpsj_66_3570_97} can all destroy the charge ordered (CO) state and lead to a ferromagnetic (FM) conducting state with a very large resistivity drop. In compounds with a CO state, the lattice is strongly coupled with spin and charge.\cite{dediu_prl_84_4489_00} When the ionic charges are ordered, local distortion changes from dynamic Jahn-Teller distortion (JTD) to collective static distortions\cite{dediu_prl_84_4489_00} and the MnO$_{6}$ octahedra buckle.\cite{yoshizawa_prb_52_13145_95} When the charge ordered insulating (COI) state is destroyed, a dramatic change in magnetic property and lattice structure occurs.\cite{kiryukhin_mrssp_494_65_98}

Under pressure, because of the bandwidth W increase, the CO state can be destroyed, and therefore a metallic state is induced. Moritomo \textit{et al}.\cite{moritomo_prb_55_7549_97} reported that pressure ($\leq$0.8 GPa) suppresses the CO state of the compound x = 0.35, 0.4, 0.5 and dT$_{CO}$/dP increases with x. In the x = 0.3 compound, pressure above 0.5 GPa induce a metallic transition which is described as a COI to ferromagnetic metallic (FMM) transition. Magnetic fields were found to be almost equivalent to pressure at least up to 1.5 GPa and can be scaled to pressure.

Because of the strong coupling between charge, spin and lattice in manganites, the investigation on the electronic and magnetic properties tuned by pressure induced lattice changes will contribute to understanding this complicated system and provide information for possible applications. At present, most of the high pressure studies on manganites were focused on the metal-insulator transition and carried out below $\sim$2 GPa. In this range, it was found that the FM state and metallic state are coupled, which can be explained qualitatively by double exchange (DE) theory.\cite{moritomo_prb_55_7549_97, neumeier_prb_52_7006_95} It is generally believed that the application of hydrostatic pressure enhances the hopping integral t$_{0}$ and leads to a systematic increase of the Curie temperature.

Pr$_{0.7}$Ca$_{0.3}$MnO$_{3}$ is charge ordered below $\sim$220 K, and is antiferromagnetic (AFM) and ferromagnetic phase-separated at low temperatures with T$_{N}$$\sim$130 K and T$_{C}$$\sim$115 K.\cite{jirak_jmmm_53_153_85} When an external pressure is applied, the low temperature COI state is destroyed and a FMM state is induced.\cite{moritomo_prb_55_7549_97, hwang_prb_52_15046_95} By considering our results at much higher pressures (up to $\sim$6.3 GPa) and other groups' results at low pressures, it was found that pressure affects the magnetic and electronic properties in a much more complicated manner than previously reported. In the range $\sim$0.8-3.5 GPa, the pressure induced metallic state at low temperature is enhanced and the metal-insulator transition (MIT) temperature (T$_{MI}$) increases with pressure. The MIT coincides with the magnetic transition in this range. Above $\sim$3.5 GPa, T$_{MI}$ decreases with pressure increase. At $\sim$5 GPa, T$_{MI}$ and magnetic transition temperature start to decouple: T$_{MI}$ drops much faster than the magnetic transition temperature and finally at $\sim$6.3 GPa the material becomes insulating and the magnetic transition temperature is almost the same as at ambient pressure.

\begin{figure}
\includegraphics[height=1.6in]{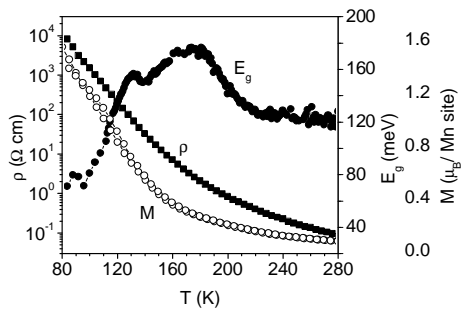}\\
\caption{\label{fig-1}Temperature dependence of resistivity, magnetization and activation energy of Pr$_{0.7}$Ca$_{0.3}$MnO$_{3}$ at ambient pressure. Magnetization (open circle) was measured in a 10 kOe magnetic field. The resistivity and the activation energy are represented by solid squares and solid circles respectively.}
\end{figure}

The samples were prepared by the solid-state reaction method. Stoichiometric amount of Pr$_{6}$O$_{11}$, MnO$_{2}$, and CaCO$_{3}$ were mixed, ground and calcined at 1100 \textcelsius\space for 70 hours; reground and calcined at 1200 \textcelsius\space for 30 hours; the powder was pressed into pellets. Then, the pellets were sintered at 1350 \textcelsius\space for 40 hours and annealed by increasing temperature to 1350 \textcelsius\space and holding it for 10 hours and then slowly cooling to room temperature at the rate of 1 \textcelsius/min. All cycles were performed in air. The x-ray diffraction pattern taken at room temperature shows that the samples are in a single crystallographic phase. The structure was refined to Pbnm symmetry using the Rietveld method. The refined lattice parameters are: a = 5.4301(1) $\mathring{\text{A}}$ {\AA}, b = 5.4676(1) $\mathring{\text{A}}${\AA}, c = 7.6751(1) $\mathring{\text{A}}${\AA}. The sample was also characterized by magnetization measurements (Fig. \ref{fig-1}). The magnetization as a function of temperature is consistent with the work of Hwang \textit{et al}.\cite{hwang_prl_75_914_95} The magnetic moment measured at 5 K is $\sim$2.1$\mu_{B}$/Mn Site, approximately equal to that reported by Jir{\'a}k \textit{et al}.\cite{jirak_jmmm_53_153_85} The details of the high pressure resistivity measurement method were described elsewhere.\cite{cui_prb_67_104107_03} Because of the lower stability of the measuring system during cooling, the data were only taken in the warming cycle.

\begin{figure}
\includegraphics[height=1.6in]{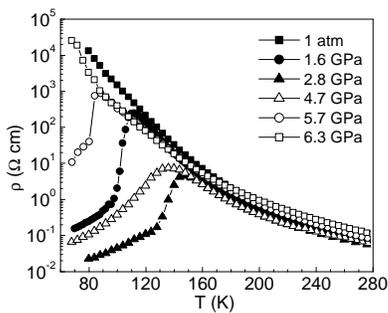}\\
\caption{\label{fig-2}Temperature dependence of resistivity of Pr$_{0.7}$Ca$_{0.3}$MnO$_{3}$ at different pressures.}
\end{figure}

\begin{figure}
\includegraphics[height=1.6in]{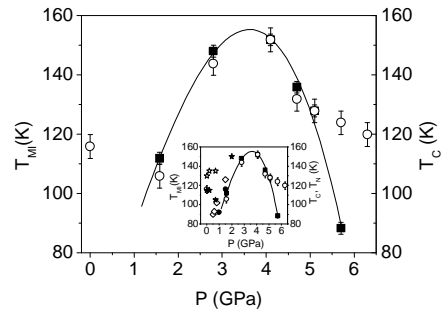}\\
\caption{\label{fig-3}Pressure induced transition temperatures. The solid squares represent T$_{MI}$, the solid line is a fit to T$_{MI}$ with a third order polynomial as a guide to the eye; the open circles represent the T$_{C}$ extracted from the activation energy. (In the inset, the results at low pressures of other authors are displayed for comparison: the solid and open stars are T$_{C}$ and T$_{N}$ estimated from the neutron scattering result in ref. \onlinecite{yoshizawa_prb_55_2729_97}; The open diamonds represent T$_{MI}$ in the warming cycle estimated from ref. \onlinecite{moritomo_prb_55_7549_97}, where the sample is single crystal; the solid circles represent T$_{MI}$ in the warming cycle estimated from ref. \onlinecite{hwang_prb_52_15046_95}, where the sample is similar to ours.)}
\end{figure}

Fig. \ref{fig-2}
 gives the temperature dependence of resistivity at different pressures. At ambient pressure, the material is insulating in the whole temperature range. As reported, pressure induces an insulator to metal transition, which is ascribed to a COI to FMM transition.\cite{moritomo_prb_55_7549_97, hwang_prl_75_914_95} With pressure increase, the transition temperature T$_{MI}$ (defined as the temperature at the resistivity peak) is shifted up while the resistivity is suppressed at the same time. In the pressure range 3$\sim$4 GPa, this trend saturates. At higher pressures, T$_{MI}$ decreases and the resistivity increases. At $\sim$6.3 GPa, the material becomes insulating in the measured temperature range and the resistivity as a function of temperature almost reproduces the case at ambient pressure. The T$_{MI}$ \textit{vs}. pressure is plotted in Fig. \ref{fig-3}. The transition temperature of our sample at low pressure is consistent to that of other authors' polycrystalline samples,\cite{hwang_prl_75_914_95} but lower than that of single crystals\cite{moritomo_prb_55_7549_97} (inset of Fig. \ref{fig-3}).

In Pr$_{1-x}$Ca$_{x}$MnO$_{3}$ (0.1$\leq$ x $\leq$0.4), the resistivity displays semiconducting behavior, with the activation energies (E$_{g}$) being slightly above 100 meV near room temperature.\cite{jirak_jmmm_53_153_85} For Pr$_{0.7}$Ca$_{0.3}$MnO$_{3}$, at ambient pressure, E$_{g}$ is $\sim$125 meV above $\sim$220 K, then increases upon cooling (Fig. \ref{fig-1}). Apparently, the E$_{g}$ increase corresponds to the charge ordering. On cooling further, E$_{g}$ decreases to $\sim$80 meV in the range of 100-130 K. It was reported that this compound is phase-separated at low temperature, with the Curie and N{\'e}el temperatures of $\sim$115 K and $\sim$130 K, respectively.\cite{jirak_jmmm_53_153_85} Hence, by comparing the temperature dependence of resistivity, magnetization and E$_{g}$ (Fig. \ref{fig-1}), the reduction of E$_{g}$ can be correlated with the magnetic transition. Yoshizawa \textit{et al}.\cite{yoshizawa_prb_55_2729_97} reported that the CO, AFM and FM transitions appear at different temperature upon cooling below 0.7 GPa, and the CO and AFM components are gradually reduced with pressure increase so that at 2 GPa only FM component is present. From the E$_{g}$ changes and comparison with the magnetization as a function of temperature and the neutron diffraction results at low pressures,\cite{yoshizawa_prb_55_2729_97} the E$_{g}$ decrease is associated with the ferromagnetic transition. Consequently, the temperature at which E$_{g }$ changes fastest with temperature is defined as T$_{C}$. The transition temperatures extracted at different pressures are shown in Fig. \ref{fig-3} together with T$_{MI}$. It is clearly seen that in the measured pressure range of $\sim$1.5-5GPa, T$_{C}$ and T$_{MI}$ coincide, indicating that pressure destroys the COI state and induces a FMM state at low temperature. But near ambient pressure and above $\sim$5 GPa, the magnetic transition and MIT are decoupled and the material becomes insulating.

In the medium pressure range, at the optimum pressure, both the magnetic transition and metal-insulator transition temperatures reach a maximum. This behavior is similar to that observed in the manganites with a larger bandwidth, in which it can be ascribed to the pressure induced Jahn-Teller distortion and Mn-O-Mn bond angle changes according to the double exchange theory.\cite{cui_prb_67_104107_03, congeduti_prl_86_1251_01, meneghini_prb_65_012111_02}

In the low ($<$$\sim$0.8 GPa) and high ($>$$\sim$5 GPa) pressure range, the material is more insulating and T$_{MI}$ and T$_{C}$ are decoupled. In thin films, it was also found that the metal-insulator transition and the magnetic transition decouple.\cite{aarts_apl_72_2975_98, rao_jap_85_4794_99} Here, the decoupling may be ascribed to the strong disorder at T$_{C}$ which can be overcome by the magnetization increasing upon cooling so that a metallic state is induced.\cite{aarts_apl_72_2975_98} The T$_{MI}$ and T$_{C}$ decoupling in bulk Pr$_{0.7}$Ba$_{0.3}$MnO$_{3}$ is ascribed to the competition between the DE and superexchange (SE) between the Mn-Mn spins.\cite{heilman_prb_65_214423_02} The SE between two neighboring Mn$^{3+}$ can either be FM or AFM depending on the Mn-Mn distance.\cite{goodenough_pr_100_564_55} The strength of SE is also a function of the bandwidth W. For the small bandwidth Pr$_{0.7}$Ca$_{0.3}$MnO$_{3}$, at low pressure because of the large lattice distortion, W is small, SE may dominate and the material displays insulating behavior. With pressure increase, due to the local distortion suppression, DE dominates and hence the insulating state is destroyed, yielding a FMM state and T$_{C}$ increases with pressure. Under pressure, the Mn-Mn distance is monotonically decreased and the local structure distortion of the MnO$_{6}$ octahedra may also change as in other manganites.\cite{cui_prb_67_104107_03, congeduti_prl_86_1251_01, meneghini_prb_65_012111_02} When it crosses over an optimum value, SE dominates again.

\begin{figure}
\includegraphics[height=1.6in]{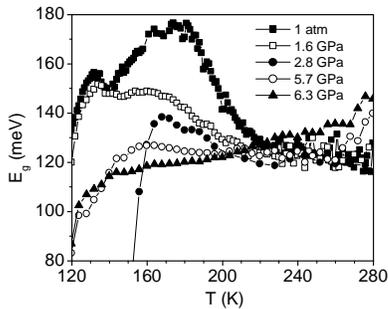}\\
\caption{\label{fig-4}Activation energy as a function of temperature at different pressures. E$_{g}$ is calculated as $dln(\rho)/d(k_{B}T)^{-1}$.}
\end{figure}

While pressure changes the electronic and magnetic states, the E$_{g}$ increase corresponding to charge ordering upon cooling disappears gradually and finally the CO state is completely suppressed so that the activation energy does not change with temperature above the magnetic transition temperature (Fig. \ref{fig-4}).
 The CO phase in the Pr$_{1-x}$Ca$_{x}$MnO$_{3}$ system is correlated with the lattice distortion and the buckling of the MnO$_{6}$ octahedra.\cite{yoshizawa_prb_52_13145_95} At T$_{CO}$, a transition from dynamic Jahn-Teller distortion to collective static distortion takes place.\cite{dediu_prl_84_4489_00} Therefore, the CO state disappearance under pressure (Fig. \ref{fig-4}) indicates that the collective static Jahn-Teller distortions are suppressed and that the dynamic distortion possibly dominates at high pressures. From the resistivity and E$_{g}$ \textit{vs}. temperature plot at $\sim$6.3 GPa, the sample seems to be a FM insulator similar to the Pr$_{0.75}$Ca$_{0.25}$MnO$_{3}$ compound.

In summary, we have measured the electrical resistivity of the small bandwidth manganite Pr$_{0.7}$Ca$_{0.3}$MnO$_{3}$ as a function of temperature and pressure (ambient to $\sim$6.5 GPa). It was found that pressure in the range above $\sim$2 GPa induces much more complicated changes than in the low pressure range in which pressure only enhances the metallic state. In the studied range, pressure does not always induce an increase in T$_{C}$ and/or T$_{MI}$, and the coupling between the magnetic transition and the metal-insulator transition is also tuned by pressure. In this small bandwidth manganite, pressure induces a metallic state in low temperature region. The metal-insulator transition temperature increases with pressure below $\sim$3.5 GPa and decreases above $\sim$3.5 GPa and the system becomes a ferromagnetic insulator at higher pressures. In the low ($<$$\sim$0.8 GPa) and high pressure regions ($>$5 GPa), the magnetic and electronic transitions are decoupled. In the medium pressure region, T$_{C}$ and T$_{MI}$ coincide and change with pressure as in other manganites with large bandwidth. The effects can be ascribed to the pressure tuned competition between DE and SE interactions. In addition to magnetic and electronic states, the charge ordering above magnetic transition temperature is suppressed completely at $\sim$3.5 GPa. High pressure measurements of the magnetic structure as well as Raman scattering measurements at different temperatures will shed light on the structural origin of these changes in the magnetic structure with pressure.  The complex behavior under pressure enables the assessment of the competition between DE and SE (which both depend on bandwidth and atomic structure) without the need for chemical doping.  Hence, this system may become a test-bed for studying highly correlated electron systems.

This work is supported by National Science Foundation Grant DMR-0209243.

\bibliography{bib-L032803}

\end{document}